\documentclass{elsart}

\usepackage{natbib}
\usepackage{amsmath, amssymb, psfrag, graphicx, tabls,xspace,setspace}
\usepackage[latin1]{inputenc}

\newcommand{\be}{\begin{eqnarray}}
\newcommand{\ee}{\end{eqnarray}}

\newcommand{\beq}{\begin{equation}}
\newcommand{\eeq}{\end{equation}}

\newcommand{\expt}[1]{\langle #1 \rangle}

\newcommand{\pavail}{p_\text{a}}

\journal{Journal of Theoretical Biology}

\begin{document}

\begin{frontmatter}

   \title{War of attrition with implicit time cost}

   \author[label1]{Anders Eriksson\corauthref{cor}}
   \ead{frtae@fy.chalmers.se}
   \author[label1]{Kristian Lindgren}
   \ead{frtkl@fy.chalmers.se}
   \author[label2]{Torbjörn Lundh}
   \ead{torbjrn@math.chalmers.se}
   \corauth[cor]{Corresponding author. Telephone: +46-31-772-3126. Fax: +46-31-772-3150}
   \address[label1]{Department of Physical Resource Theory, Chalmers University of Technology and Göteborg University, SE-41296 Göteborg, Sweden}
   \address[label2]{Department of Mathematics, Chalmers University of Technology and Göteborg University, SE-41296 Göteborg, Sweden}

\begin{abstract}
%\singlespacing
\raggedright

In the game-theoretic model \emph{war of attrition}, players are
subject to an explicit cost proportional to the duration of
contests. We construct a model where the time cost is not
explicitly given, but instead depends implicitly on the strategies
of the whole population. We identify and analyse the underlying
mechanisms responsible for the implicit time cost. Each player
participates in a series of games, where those prepared to wait
longer win with higher certainty but play less frequently. The
model is characterised by the ratio of the winner's score to the
loser's score, in a single game. The fitness of a player is
determined by the accumulated score from the games played during a
generation. We derive the stationary distribution of strategies
under the replicator dynamics. When the score ratio is high, we
find that the stationary distribution is unstable, with respect to
both evolutionary and dynamical stability, and the dynamics
converge to a limit cycle. When the ratio is low, the dynamics
converge to the stationary distribution. For an intermediate
interval of the ratio, the distribution is dynamically but not
evolutionarily stable. Finally, the implications of our results
for previous models based on the war of attrition are discussed.
\end{abstract}


\end{frontmatter}

\raggedright

% --------------------------------------------------------------
% --------------------------------------------------------------
% --------------------------------------------------------------
% --------------------------------------------------------------

\section{Introduction} \label{sec:introduction}

In many interactions between individuals, the time that passes
from the start to the separation may be of importance for how to
evaluate the outcome for the participants. Most game-theoretic
situations do not take this into account, but it is generally
assumed that the time passed is independent of the strategies. In
the war of  attrition, originally introduced by
\citet{maynard-smith_price73}, the length of the interaction is
explicitly accounted for by a cost directly affecting the score
for the players. Our starting-point will be the situation
described and motivated in the introduction to Chapter 3 in
\citep{maynard-smith82}. This is a waiting-game for two players,
where the one who gives up and quits gets a smaller reward, the
\emph{consolation prize} $k < 1$ compared to the score 1 for the
one who stays, but both pay a cost $c\,t$, where $t$ is the
duration of the game and $c$ is a positive constant. The cost does
not necessarily come from the contest itself; the engagement in a
contest may take time from possible alternative activities, for
example, there is less time to gather resources for survival and
reproduction. A strategy is a certain \emph{waiting-time} that the
player is prepared to wait unless the opponent finishes before.
When two players meet the one with the largest waiting-time wins,
and in case of a draw both get $(1 + k)/2$.

In the game of attrition, as stated above, there is a mixed
strategy defining a Nash equilibrium for the game, given by the
probability density $p(x) = c/(1-k)\,e^{-c\,x/(1-k)}$, where $x$
is the waiting-time. It should be noted though, that Nash
equilibria are also given by strategy pairs where one has
waiting-time 0 and the other waits long enough, for example by
always waiting at least $(1-k)/c$. These are the only types of
Nash equilibria that can exist in this game.

If the game is put into a co-evolutionary context, using
replicator dynamics, it is known that the mixed strategy Nash
equilibrium corresponds to an evolutionarily stable population
mixture of pure strategies \citep{maynard-smith74}. In this case,
with an explicit time cost, the consolation prize $k$ is not
critical since the fitness can be transformed to the case with $k
= 0$ by an affine transformation (multiplication with a positive
constant and subtraction by any constant), that does not affect
the evolutionary dynamics of the population.
\citet{bishop_cannings78b} study generalised score functions for
this game.

It is clear that if there is no cost associated with the duration
of a contest, waiting forever is a Nash equilibrium. But if we
study this game in a co-evolutionary context, we may implicitly
include a cost of time by letting those who are involved a shorter
time in the games participate in a larger number of games,
allowing them to accumulate a higher score. In this case, the
consolation prize $k$ for the loser plays a crucial role.

In this paper, we investigate the characteristics of the
co-evolutionary dynamics of a population interacting according to
the game of attrition with no explicit time cost. The players in
the population can be in one of two states: either they are
involved in a contest with another player, or they are available
for entering a new contest. The activity of the players in the
population, during a generation, is modelled as a process that
randomly selects pairs of available players to engage in contests.
This leads to an implicit time cost, which is higher for players
involved in longer games. In particular we investigate the
existence of stationary distributions and how their stability
depends on the consolation prize $k$, the only free parameter in
the model.

The war of attrition has been extended in various ways, including
multi-person games \citep{haigh_cannings88} and generalisations of
the payoff structure \citep{bishop_cannings78b}.

\citet{cannings_whittaker95} studied a modification of the model
by \citet{maynard-smith82} similar to the one we present. They
suggest a mechanism that implies more games for players that
finish faster, but keep the explicit time cost. Unlike our model,
their approach is restricted to positive integer waiting-times.
The time between games is always one unit of time, the same as the
smallest possible waiting-time. In Section~\ref{sec:comparison to
other models} their model is discussed in more detail. Other
variants of their model are presented in
\citep{cannings_whittaker94, whittaker96}, where a fixed amount of
resource is to be divided between a number of contests based on
the war of attrition.

An earlier study, by \citet{hines77,hines78} also takes into
account that strategies determine the number of games played.
Hines analyses a model where animals forage for food. When an
animal finds a piece of food, with a given probability it may
consume the food undisturbed, otherwise it enters a war of
attrition for the food parcel. The details of the model is
described in Section~\ref{sec:comparison to other models}, where
we also discuss the implications of our results for the model by
Hines.

The paper is organised as follows: In Section~\ref{sec:social
dynamics model}, we describe the details of the social dynamics
model that governs how pairs of individuals engage in games. In
Section~\ref{sec:evolutionary dynamics}, we present how the
long-term changes in the composition of strategies in the
population is governed by the replicator dynamics. In Sections
\ref{sec:deterministic strategies} and \ref{sec:stationary
distributions}, the stationary distribution for deterministic
strategies under the replicator dynamics is derived. The stability
analysis of the stationary distribution is presented in
Section~\ref{sec:dynamic stability}, with details shown in
Appendix~\ref{app:Dynamic stability details}. In
Section~\ref{sec:global_convergence} we study the long-term
evolution of the dynamics, from different initial distributions.
In Section~\ref{sec:explicit}, we show how models with an explicit
time cost can be mapped to our model. In
Section~\ref{sec:comparison to other models} we relate or model to
the models of Hines and Cannings and Whittaker. We conclude with
summarising and discussing our results in
Section~\ref{sec:discussion}.

% --------------------------------------------------------------
% --------------------------------------------------------------
% --------------------------------------------------------------
% --------------------------------------------------------------

\section{Social dynamics model} \label{sec:social dynamics model}

\begin{figure}
   \centerline{\includegraphics{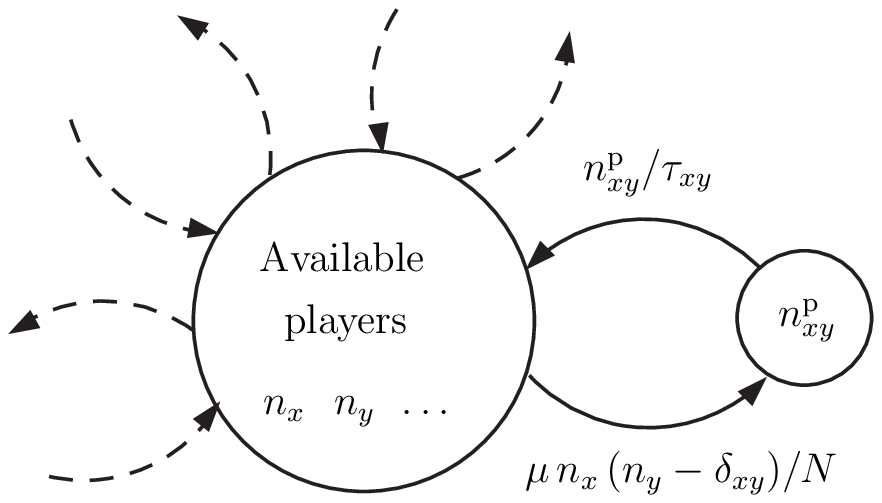}}
\caption{\label{fig:social_dynamics}
An illustration of the processes in the social dynamics. Players
with strategies $x$ and $y$ meet and engage in contests. When a
contest ends, the participants return to the available state. The
number of individuals of strategy $x$ in the available state is
$n_x$, and the number of ongoing contests between players with
these strategies is given by $n^\text{p}_{xy}$, for a given pair
of strategies $(x,y)$. The rates at which contests start and end
are given along the arrows, where $\tau_{xy} $ is the expected
duration of a $(x,y)$-game, and $N$ is the whole population size.
The dotted arrows connect to the nodes for other pairs.
}
\end{figure}

Consider a population of $N$ individuals. During the course of one
generation, an individual experiences a series of encounters with
other individuals. It is assumed that a fraction of the encounters
lead to a conflict, from competing interests, which is resolved
through a contest. We assume that no more than two players meet in
the same contest, so that only pair-wise interactions are
considered. Thus, the state of the whole population is controlled
by two processes: one in which available players form pairs and
become engaged in contests, and one in which pairs break up and
make players available. See Fig.~\ref{fig:social_dynamics} for an
illustration of these processes. We denote the combined processes
the \emph{social dynamics} of the population.

% fig 1
%======================================================================
%======================================================================

A contest between two individuals, with waiting-times $x$ and $y$,
takes the form of the war of attrition game. In general, $x$ and
$y$ may be stochastic variables, reflecting mixed strategies. From
Section~\ref{sec:deterministic strategies} and onward, we focus on
pure strategies. The duration of the game is given by the smallest
of the waiting-times. The player with the largest waiting-time
gets the score 1, and the other player gets the consolation prize,
i.e., the score $k$, with $0 < k < 1$. If both players have the
same waiting-time, they both get an expected score of $(1+k)/2$.

At each instant a fraction $\pavail$ of the population is
available for playing, while the rest of the individuals are
engaged in pair-wise contests. The total number of players with
strategy $x$ in the population is denoted by $n^\text{tot}_x$, the
number of available players with waiting-time $x$ is given by
$n_x$, and the number of ongoing games between the players with
waiting-times $x$ and $y$ is $n^\text{p}_{xy}$. It is convenient
to treat the pairs $(x,y)$ and $(y,x)$ separately; though they are
equivalent, it simplifies the calculations.

We assume that the time intervals between games are independent.
Available players form ordered pairs $(x, y)$ for playing
according to a Poisson process with rate $\mu(\pavail)\,n_x\,(n_y
- \delta_{xy})/N$, where $\delta_{xy}$ equals one if $x = y$ and
zero otherwise. Note that time is measured in units of $N$ for
convenience: the number of events per unit of time remain finite
for large $N$. The rate $\mu(\pavail)$ is a positive function of
the fraction of available players, and may be used to model how
the social dynamics depends on the availability of players. For a
player, the expected time between the end of one contest and the
onset of the next is $[2\,\pavail\,\mu(\pavail)]^{-1}$. We
illustrate the choice of $\mu(\pavail)$ by two examples. First, a
model where players perform random walks, and occasionally meet
and play. This corresponds to constant $\mu(\pavail)$. Second, a
model where the expected time between games for a player is
independent of the number of available players, corresponds to
taking $\mu(\pavail) \propto \pavail^{-1}$. In general, we assume
that $\mu(\pavail)$ is chosen so that the expected time between
games for a player is decreasing with $\pavail$ or constant, and
is bounded for $0 \le \pavail \le 1$.

When the population is finite, the social dynamics forms a Markov
process and we know that there is a unique stationary distribution
characterising the distribution of strategies in the different
states.

In general, we should expect that players who are prepared to wait
for a longer time (before finishing a game) enter the available
state less frequently, and it is this effect that will result in
an implicit time cost. We can find the equilibrium distribution,
i.e., the expectation value of the number of individuals and
pairs, by using the requirement that the expected number of pairs
formed equals the number of pairs that finish playing, per unit of
time:
\beq\label{eq:balance}
   \frac{\expt{n^\text{p}_{xy}}}{\tau_{xy}} = \frac{\mu(\pavail)}{N} \expt{n_x} \left(\expt{n_y} - \delta_{xy}\right),
\eeq
where $\tau_{xy}$ is the expected duration of a game between a
pair of players with strategies $x$ and $y$, respectively. By
counting players with strategy $x$ that are available or playing,
we get the relation
\beq\label{eq:ntot_balance}
   n^\text{tot}_x \ =\ \expt{n_x} + 2 \sum_y
   \expt{n^\text{p}_{xy}} \ =\ \expt{n_x} + 2\, \frac{\mu(\pavail)}{N} \sum_y \tau_{xy}
   \expt{n_x} \left( \expt{n_y} - \delta_{xy}\right),
\eeq
which can be used to solve for the equilibrium distribution for
players in the available state, given a certain overall
distribution of strategies.

We now focus on large populations. In the limit of large $N$, we
take $u(x)$ to be the distribution of the waiting-time $x$ in the
population ($u(x) \sim n^\text{tot}_x /N$), and $\rho(x)$ to be
the distribution of $x$ among the available players ($\rho(x) \sim
n_x / \sum_y n_y$). In the limit of many players it is convenient
to consider a general distribution of waiting-times. Thus,
(\ref{eq:ntot_balance}) becomes
\beq\label{eq:equilibrium}
    u(x) = \pavail\,\rho(x) + 2\,\pavail^2 \mu(\pavail) \int_0^\infty  \tau(x,y)\, \rho(x)\, \rho(y)  \,dy,
\eeq
where $\tau(x,y)$ is the expected duration of a game between
players with strategies $x$ and $y$. Note that different strategy
spaces may be modelled by the appropriate function $\tau(x,y)$ of
$x$ and $y$, where now $x$ (and $y$) is a parameter characterising
the strategy rather than the explicit waiting-time. For instance,
the set of mixed strategies with exponentially distributed
waiting-times may be characterised by the expected waiting-time.

Since the expected time between contests for a player is
$[2\,\pavail\,\mu(\pavail)]^{-1}$, the expected number of games
per unit of time is
\beq\label{eq:N_G(x)}
   N_\text{G}(x) = \left[\, \frac{1}{2\,\pavail\,\mu(\pavail)} + \int_0^\infty \tau(x,y)\,\rho(y) \,dy \,\right]^{-1}.
\eeq
We may now express $u(x)$ as
\beq\label{eq:u(x)}
   u(x) = 2\,\pavail^2\,\mu(\pavail) \, \rho(x) / N_\text{G}(x).
\eeq

As an alternative, we may derive (\ref{eq:u(x)}) from the
ergodicity of the system. The fraction of the population with
parameter $x$ that is in the available state is
$\pavail\,\rho(x)/u(x)$. A player with parameter $x$ spend a
fraction $N_\text{G}(x)/[2\,\pavail\,\mu(\pavail)]$ of the time in
the available state. Since the onsets of two consecutive contests
are assumed to be uncorrelated, the two fractions must be equal,
and we again arrive at (\ref{eq:u(x)}).

% --------------------------------------------------------------
% --------------------------------------------------------------
% --------------------------------------------------------------
% --------------------------------------------------------------

\section{Evolutionary dynamics} \label{sec:evolutionary dynamics}

It is assumed that the population evolves so that players with
strategies that get higher scores per unit time increase their
fraction of the population, leading to a change in the
distribution $u(x,t)$ at time $t$. Further, it is assumed that the
evolutionary dynamics and the social dynamics occur on separate
time scales so that the equilibrium distribution given by
(\ref{eq:equilibrium}) can be used to determine the scores of the
different players. Then, on the evolutionary time scale, the
population dynamics is given by the replicator dynamics,
\beq\label{eq:evolutionary dynamics}
    \frac{\partial u(x,t)}{\partial t}  = u(x,t) \left[\, f(x,t) - {\bar{f}(t)}\, \right],
\eeq
where $f(x,t)$ is the expected fitness of a player with strategy
$x$, per unit of time in the social dynamics, and $\bar{f}(t)$ is
the average fitness in the whole population at time $t$,
\[
    \bar{f}(t) = \int_0^\infty u(x,t)\, f(x,t) \, dx.
\]
To simplify the notation, let us suppress the explicit time
dependence in the notation and write $u(x)$ for $u(x,t)$ etc. The
expected fitness of a player with strategy $x$ is the product of
two factors, the expected score per game and the number of games
per unit of time:
\beq\label{eq:payoff}
    f(x) =  N_\text{G}(x) \int_0^\infty \left[\, k + (1 - k)\,P(x,y) \,\right] \, \rho(y) \, dy ,
\eeq
where $P(x,y)$ is the probability that a player with strategy $x$
is the winner in a game against a player with strategy $y$.

A player can increase the expected score per game by waiting
longer, but that will decrease the number of games played per unit
of time, and we get an implicit cost due to the longer engagement
in the games. Therefore, whether it is advantageous for a player
to increase or decrease the waiting-time depends on the
distribution of strategies in the population.

Note that the sum of the scores in each game is $1+k$, and the
total number of games per unit of time is
$2\pavail^2\mu(\pavail)$. From this we see a general connection
between average fitness and the fraction of available players:
\begin{lem}\label{thm:avg_payoff}
When the population is in equilibrium with respect to the social
dynamics, the average fitness in the population is
\beq \label{eq:fbar}
    \bar{f} = \pavail^2 \, \mu(\pavail)\, (1+k),
\eeq
for any set of strategies.
\end{lem}

% --------------------------------------------------------------
% --------------------------------------------------------------
% --------------------------------------------------------------
% --------------------------------------------------------------

\section{ Deterministic strategies }
\label{sec:deterministic strategies}

The formalism presented in the previous sections applies to both
pure and mixed strategies. In the following, we assume pure
strategies, so that each player is prepared to wait a given time
$x$, and the population is then characterised by a distribution
$u(x)$ over the waiting-times. When a player $A$ with strategy $x$
meets a player $B$ with strategy $y$, the probability of winning
for player $A$ is $P(x,y) = \theta(x - y)$, where $\theta(z)$ is
one if $z > 0$, $1/2$ if $z = 0$ and zero otherwise. The duration
of the game is given by $\tau(x,y) = \min(x,y)$.

When there is an atom at infinity (i.e., a finite share of the
population with infinite waiting-time), the players that wait
indefinitely always get a fitness of zero, since the expected
score per game is bounded from above by one, and the expected time
per game is infinite (c.f.\@ (\ref{eq:payoff})). Thus, this
strategy is dominated by all other strategies, and in the
following we assume there is no atom at infinity.

\begin{figure}
   \centerline{\includegraphics{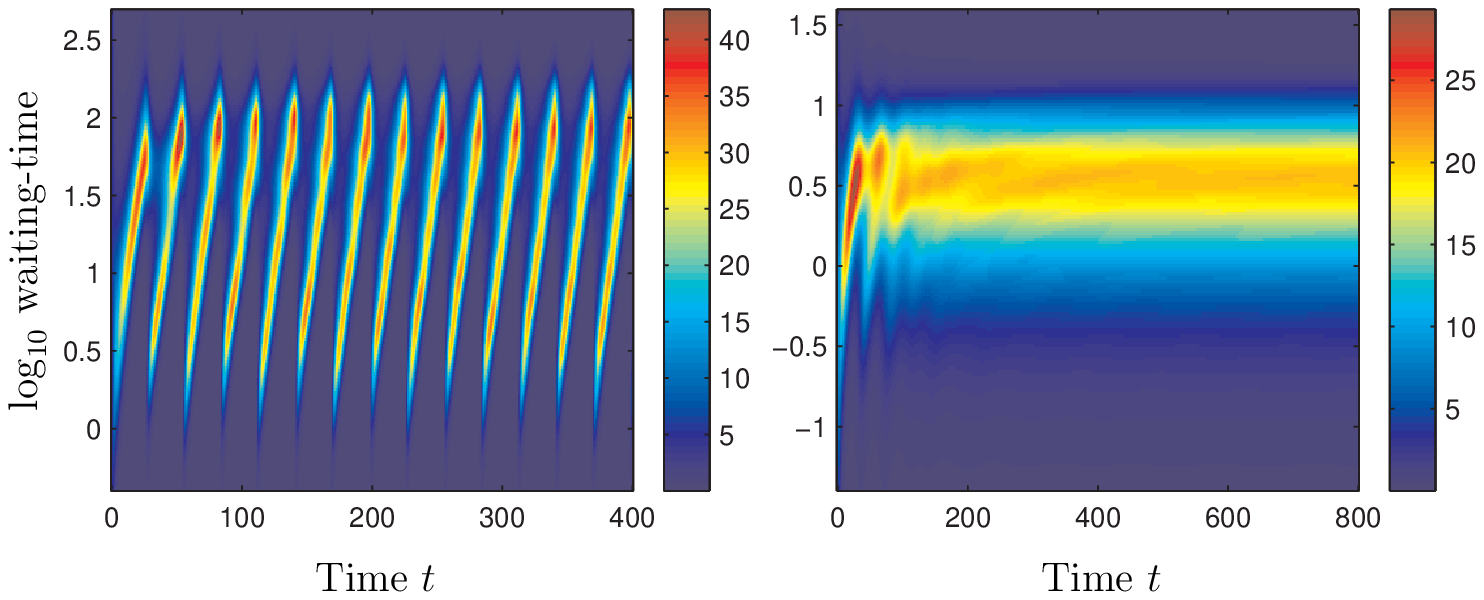}}
\caption{\label{fig:time_evol}
The time evolution of the distribution of waiting-times in the
population, in the case of constant $\mu(\pavail)$, for
consolation price $k = 0.1$ (left) and $k = 0.3$ (right). For
clarity, we show the distribution of the base-10 logarithm of the
waiting-time in the population. The colour-bars show the value
corresponding to each colour in the figures. For $k = 0.1$, the
dynamics seems to converge to a stable limit cycle, and for $k =
0.3$ it seems to converge to a stationary distribution.
}
\end{figure}

In Fig.~\ref{fig:time_evol}, we show the time evolution of the
distribution of waiting-times in the population, in the case of
constant $\mu(\pavail)$, for consolation price $k = 0.1$ and $k =
0.3$, from numerical integration of the replicator dynamics (c.f.
(\ref{eq:evolutionary dynamics})). The initial distribution is
taken to be proportional to the inverse waiting-time over the
interval shown in the figure, and zero elsewhere. For $k = 0.1$,
we find that the dynamics seems to converge to a limit cycle, and
for $k = 0.3$ it seems to converge to a stationary distribution.

% fig 2
%======================================================================
%======================================================================

% --------------------------------------------------------------
% --------------------------------------------------------------
% --------------------------------------------------------------
% --------------------------------------------------------------

\section{Stationary distributions}
\label{sec:stationary distributions}

The numerical simulations show that the systems seems to converge
to a stationary distribution for some values of the consolation
prize $k$. In this section, we derive the form of this
distribution.

A stationary distribution $u(x)$ is determined by the requirement
$f(x) = \bar{f}$ wherever $u(x) > 0$. With the cumulative
distribution of the available players, $\Phi(x) = \int_0^x\!
\rho(y) \, dy$, we can use (\ref{eq:N_G(x)}) and (\ref{eq:payoff})
to express the stationarity condition as
\beq\label{eq:stationarity_condition}
   \Phi(x) + k \big(1 - \Phi(x)\big) =
   \bar{f} \, \left[ \frac{1}{2 \, \pavail \, \mu(\pavail)} + \int_0^x y \,\rho(y) \, dy + x \big(1 - \Phi(x)\big) \right].
\eeq
Next, we take the derivative of this equation with respect to $x$:
\[
  (1 - k)\,\rho(x) =  \bar{f} \, [\, x\, \rho(x) - x\, \rho(x) + 1 - \Phi(x) \,] =  \bar{f} \, [\,1 - \Phi(x)\,] .
\]
Since $\Phi(0) = 0$ and $\Phi(\infty) = 1$, we see that there is a
unique solution
\beq\label{eq:det_solution_v}
    \Phi(x) = 1-e^{-\lambda x} \text{ where } \lambda = \frac{\bar{f}}{1 - k} = \pavail^2 \, \mu(\pavail)\,\frac{1+k}{1-k}
\eeq
by Lemma \ref{thm:avg_payoff}. By inserting the solution
(\ref{eq:det_solution_v}) into (\ref{eq:stationarity_condition})
we find the fraction of available players in the population:
\[
   \pavail = \frac{2\,k}{1+k}.
\]
Given the distribution of strategies among the available players,
the distribution $u(x)$ of strategies in the population can be
calculated from (\ref{eq:equilibrium}). The results are summarised
in the following theorem:
\begin{thm} \label{thm:distribution}
In a population of players with deterministic waiting-times, there
is a unique stationary distribution $u(x)$ involving all
strategies $x$, given by
\beq\label{eq:det_stationary_distr}
    u(x) = \frac{2\, \lambda}{1+k} \left[\, e^{- \lambda\, x} - (1-k)\, e^{- 2\,\lambda\, x} \,\right] \text{, where }  \lambda \ =\ \frac{4\,k^2}{1 - k^2} \,\mu(\frac{2\,k}{1+k}).
\eeq
The fraction of available players is then $\pavail = 2\, k/(1 +
k)$ and the expected fitness of all players is
\[
   \bar{f} = \frac{4\, k^2}{1 + k} \,  \mu\big(\frac{2\,k}{1+k}\big) .
\]
Furthermore, the distribution of strategies among the available
players is exponentially distributed:
\[
   \rho(x) = \lambda\,e^{-\lambda\, x}.
\]
\end{thm}

From Theorem~\ref{thm:distribution}, we obtain the distribution of
strategies amongst the players engaged in games at the stationary
distribution as
\[
   \frac{u(x) - \pavail\,\rho(x)}{1 - \pavail} = 2 \lambda e^{-\lambda x} - 2 \lambda e^{-2\lambda x}.
\]

\begin{figure}
   \centerline{\includegraphics{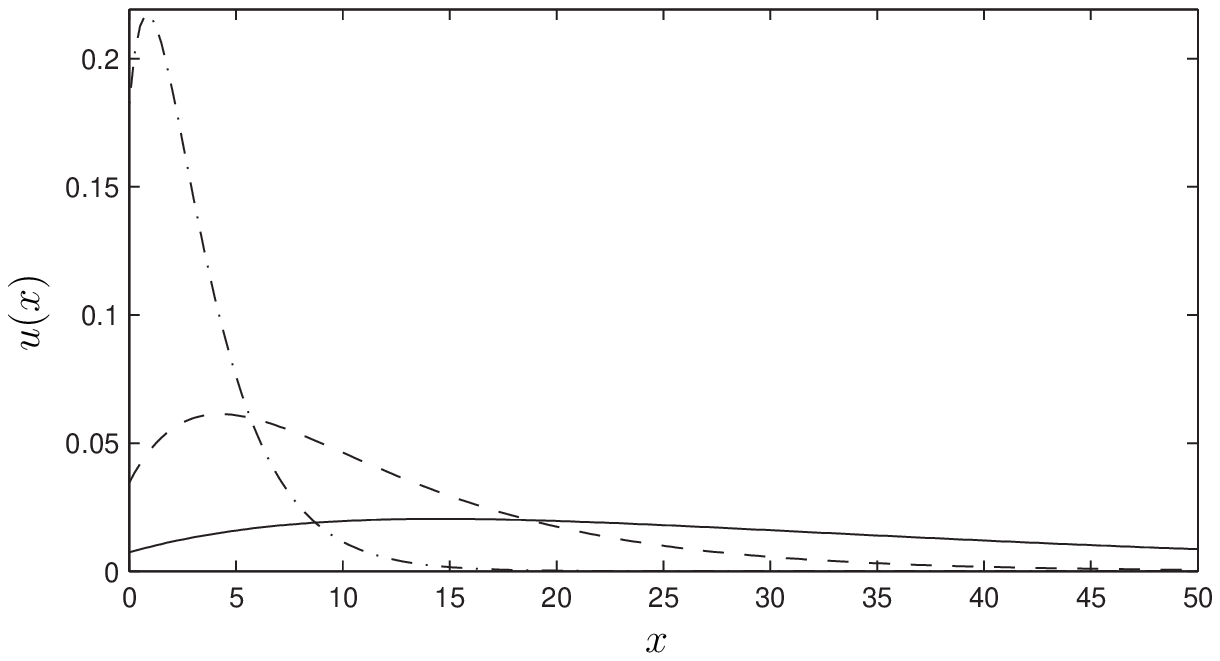}}
\caption{\label{fig:stat_distr} The stationary distribution $u(x)$
for the consolation prize  $k = 0.1$ (solid line), $k = 0.17$
(dashed line) and $k = 0.3$ (dashed and dotted line).}
\end{figure}

The stationary distribution is shown in Fig.~\ref{fig:stat_distr}
for three values of the consolation prize $k$. The strategies of
the available players are exponentially distributed, as is the
stationary distribution in the original war of attrition
\citep{maynard-smith74}. In our model, however, the stationary
distribution is decreasing for short waiting-times. This is due to
that strategies with short waiting-time are over-represented among
the available players, compared to the distribution of strategies
in the population. Of the players with waiting-time $x$, a
fraction $\pavail\,\rho(x)/u(x) = k/[1 - (1-k)\exp(-\lambda x)]$
is in the available state, at the stationary distribution. This
means that, despite the risk for getting stuck in long games,
players with long waiting-times will spend at least a fraction $k$
of their time in the available state.

% fig 3
%======================================================================
%======================================================================

% --------------------------------------------------------------
% --------------------------------------------------------------
% --------------------------------------------------------------
% --------------------------------------------------------------

\section{Stability of the stationary distribution}
\label{sec:dynamic stability}

Although there exist stationary distributions for all $k\in (0,1)$
as seen in Fig.~\ref{fig:stat_distr}, we note from
Fig.~\ref{fig:time_evol} that there could be a fundamental
difference in the dynamics for different consolation prizes $k$.
In order to better understand the dynamics of the model, we
investigate the stability of the stationary distribution
(\ref{eq:det_stationary_distr}), as function of the consolation
prize $k$. We give analytical results for infinitesimal
perturbations of the stationary distribution, as well as results
from numerical simulations, in order to explore the global
convergence to the stationary distribution.

Both the stationary distribution $u(x)$ and the distribution of
strategies among available players, $\rho(x)$, are polynomials in
$e^{-\lambda\,x}$. In order to simplify the ana\-ly\-sis of the
stability of the stationary distribution, and also to facilitate
more accurate simulations, we map the strategy $x$ to a point $s$
in the unit interval through
\[
   s = e^{-\lambda x} \quad \text{with} \quad \lambda = \frac{4\,k^2}{1 - k^2} \, \mu\big(\frac{2\,k}{1+k}\big).
\]
Thus the whole range of waiting-times $[0,\infty)$ is represented
by the interval $(0,1]$. The distribution of $s$ in the population
is then $u_s(s) = u(-\lambda^{-1}\,\ln s) / (\lambda s)$.
Similarly, we define $f_s(s)$ as the fitness of a player with
parameter $s$.

In order to be specific, in the following we restrict the analysis
of the stability properties to the case where $\mu(\pavail)$ is
constant. It is straight-forward to extend the analysis to other
$\mu(\pavail)$. With the functions
\be
   g(s) &=& \frac{1+k}{2\,k}\,\frac{1}{\lambda\,s}\, \pavail \, \rho(-\frac{\ln s}{\lambda}) \quad \text{and}\nonumber\\
   h(s) &=& k + (1-k)\,\lambda \int_0^1  \tau(\,-\frac{\ln s}{\lambda}, -\frac{\ln t}{\lambda})\, g(t)\,dt, \nonumber
\ee
we can express (\ref{eq:u(x)}) as
\be \label{eq:express}
   u_s(s) = 2\,g(s)\,h(s)/(1+k).
\ee
Note that $g(s)$ is proportional to the distribution of $s$ among
the available players. For convenience, we now measure time in
units of $\bar{f}$ in the stationary distribution, $4k^2/(1+k)$.
In the new units, we have $f_s(s) = m(s)/h(s)$ and $\bar{f} =
\hat{g}^2$, where
\[
   m(s) = \int_0^1 S(-\frac{\ln s}{\lambda}, -\frac{\ln t}{\lambda})\,g(t)\,dt
   \quad\text{and}\quad \hat{g} = \int_0^1 g(t)\,dt.
\]
The time evolution of the distribution of $s$,
(\ref{eq:evolutionary dynamics}),
 is then
\[
   \frac{\partial u_s(s)}{\partial t} = \frac{2}{1+k}\, g(s)\, [\, m(s) - \hat{g}^2\, h(s)\,] .
\]
For deterministic strategies, we have
\be
   \tau(-\frac{\ln s}{\lambda},-\frac{\ln t}{\lambda})  &=& - \frac{1}{\lambda} \max(\,\ln s, \ln t\,) \nonumber\\
   S(-\frac{\ln s}{\lambda}, -\frac{\ln t}{\lambda}) &=&  \theta(t-s) + k\,\theta(s-t),\nonumber
\ee
so at the stationary distribution (\ref{eq:det_stationary_distr})
we get $g(s) = \hat{g} = 1$ and $\ h(s) = m(s) =  1 - (1-k)s$.
Thus, in the parameter $s$ the stationary distribution becomes
\beq \label{eq:losning_stat}
  u_s(s) = 2\,\left[ 1 - (1-k)s \right]/(1+k).
\eeq

% --------------------------------------------------------------
% --------------------------------------------------------------
% --------------------------------------------------------------
% --------------------------------------------------------------

\subsection{Evolutionary stability}

There are many different ways of characterising the stability of a
distribution under the evolutionary dynamics in the literature
(see e.g. \citet{hines87} for a review). One of the most commonly
used definitions of stability is to test whether the stationary
distribution can resist an invasion from a population with
distribution $q_s(s)$ at a level $\eta$, for all positive values
of $\eta$ in a neighbourhood of zero. The distribution of
strategies in the population after the invasion is thus $(1 -
\eta)\,u_s(s) + \eta\,q_s(s)$. Then $u_s(s)$ is an evolutionarily
stable distribution if and if only the expected fitness among the
players in the old distribution is higher than the fitness among
the invaders:
\[
   \int_0^1 u_s(s)\,\tilde{f}_s(s) \,ds > \int_0^1 q_s(s)\,\tilde{f}_s(s) \,ds ,
\]
where $\tilde{f}_s(s)$ is the fitness of a player with strategy
$s$ in the population with distribution $\tilde{u}_s(s)$. Now
expand $\tilde{f}_s(s)$ for small $\eta$: $ \tilde{f}_s(s) =
f_s(s) + \delta f_s(s)\,\eta + O(\eta^2)$, where $\delta f_s(s) =
\mathcal{F}[\,\delta u_s(s)\,]$ is a functional of the
perturbation $\delta u_s(s)=\tilde{u}_s(s)-u_s(s)$. Since $f_s(s)
= \bar{f}$ at the stationary distribution,
\[
   \int_0^1 [\,u_s(s) - q_s(s)\,]\,f_s(s) \,ds = \int_0^1 [\,u_s(s) - q_s(s)\,]\,\bar{f} \,ds \ =\ 0,
\]
so $u_s(s)$ is evolutionarily stable if and only if for all small
enough perturbations $\delta u_s(s)$,
\[
    \int_0^1 \delta u_s(s) \, \mathcal{F}[\,\delta u_s(s)\,] \,ds < 0 .
\]
Since we know how to calculate the distribution $u_s(s)$ from the
distribution of available players, it is convenient to express a
perturbation $\delta u_s(s)$ as a perturbation in the availability
of the strategies using (\ref{eq:express}):
\[
   \delta u_s(s) = \frac{2}{1 + k} \mathcal{L}[\,\delta g(s)\,] \text{ where } \mathcal{L}[\,\delta g(s)\,] = \delta[\, g(s) h(s) \,] .
\]
At the stationary distribution, we get
\[
   \mathcal{L}[\,\delta g(s)\,] = [\, 1 - (1-k)s \,] \, \delta g(s) - (1-k) \, \int_0^1 \max(\ln s,\ln t) \, \delta g(t) \, dt  .
\]
Since $\delta u_s(s)$ is normalised to zero, we have that the
corresponding perturbation $\delta g(s)$ must obey
\beq\label{eq:dg norm cond}
  0 = \int_0^1 \mathcal{L}[\delta g(s)] \,ds
    = \int_0^1 [\, 2 - k - 2\,(1-k)\,s \,] \, \delta g(s) \,ds .
\eeq
With $\delta f_s(s) = \mathcal{B}[\,\delta g(s)\,]$, we find
\[
   \mathcal{B}[\,\delta g(s)\,] =
   \delta [\, \frac{m(s)}{h(s)} \,] =
   \frac{\delta m(s)\, h(s) - m(s)\,\delta h(s)}{h(s)^2} = \frac{\delta m(s) - \delta h(s)}{1 - (1-k)\,s}
\]
at the stationary distribution (\ref{eq:losning_stat}), where
\[
   \delta m(s) - \delta h(s) = \int_0^s ds\,[\, k + (1-k)\,\ln\, s \,]\,\delta g(t) +
          \int_s^1 ds\,[\, 1 + (1-k)\,\ln\, t \,]\,\delta g(t) .
\]
Thus $u_s(s)$ is not evolutionarily stable if we can find a
perturbation $\delta g(s)$ such that
\[
  \mathcal{M}[\,\delta g(s)\,] = \int_0^1 \mathcal{L}[\,\delta g(s)\,] \mathcal{B}[\,\delta g(s)\,] \,ds > 0.
\]
We expand the perturbation $\delta g(s)$ in the shifted Legendre
polynomials $\bar{P}_i(s)$ \citep{abramowitz_stegun72}, and
maximise $\mathcal{M}[\,\delta g(s)\,]$ subject to the
normalisation condition (\ref{eq:dg norm cond}) and $\int_0^1
\mathcal{L}[\,\delta g(s)\,]^2 \, ds = 1$. In order to find a
numerical solution, we restrict the analysis to the $n+1$ first
Legendre polynomials: $\delta g(s) = \sum_{i=0}^{n} g_i \,
\bar{P}_i(s)$. Ignoring the higher order polynomials amounts to
ignoring highly oscillatory contributions to the perturbation
$\delta g(s)$.

With $n = 50$ we find that for $k < k^*_\text{ess} \approx 0.5196$
there is a positive maximum of $\mathcal{M}[\,\delta g(s)\,]$, but
when $k > k^*_\text{ess}$ all extreme values are negative. Thus we
have evidence that there is no evolutionarily stable strategy for
$k < k^*_\text{ess}$. For $k > k^*_\text{ess}$ we have not shown
that the distribution is evolutionarily stable, but that it is
likely so. The value $k^*_\text{ess}$ seems not to be sensitive to
the number of Legendre polynomials $n$ used in the expansion of
$\delta g(s)$, when $n \gtrsim 30$.

We have also performed these calculations for the case
$\mu(\pavail) \propto \pavail^{-1}$. For $n = 50$ we find
$k^*_\text{ess} \approx 0.6738$, and again the value of
$k^*_\text{ess}$ seems to converge when $n$ is large enough (i.e.
$n \gtrsim 30$).

Note that, in Fig.~\ref{fig:time_evol}, the distribution seems to
converge to the stationary distribution for $k$ as low as $0.3$.
This indicates that evolutionary stability of the stationary
distribution implies dynamical stability, but that the converse
does not hold. In the following section, we address this issue.

% --------------------------------------------------------------
% --------------------------------------------------------------
% --------------------------------------------------------------
% --------------------------------------------------------------

\subsection{Dynamical stability}

We examine the long-term response of the replicator dynamics to a
small perturbation $\delta u_s(s)$ in the composition of the
population, around the stationary solution:
\beq\label{eq:perturbed_evol}
   \frac{\partial \delta u_s(s)}{\partial t} = \frac{2}{1+k}\, \delta[\, g(s)\,m(s) - \hat{g}^2\,g(s)\,h(s)\,] .
\eeq
With the linear operator
   $\mathcal{A}[\delta g(s)] = \delta[\, g(s)\,m(s) - \hat{g}^2\,g(s)\,h(s) \,]$,
we may write (\ref{eq:perturbed_evol}) in terms of the time
evolution of the corresponding perturbation $\delta g(s)$ as:
\beq\label{eq:perturb_lin_op}
   \frac{\partial}{\partial t} \mathcal{L}[\delta g(s)] = \mathcal{A}[\delta g(s)].
\eeq
Note that the perturbation $\delta g(s)$ is subject to the
constraint (\ref{eq:dg norm cond}). See Appendix~\ref{app:Dynamic
stability details} for the details of $\mathcal{A}$. In order to
find whether a perturbation $\delta g(s)$ grows or decays, we look
for solutions to (\ref{eq:perturb_lin_op}) on the form
   $\delta g(s) = \phi(s)\,e^{\gamma t}$,
where $t$ is the evolutionary time. The perturbation $\delta g(s)$
grows if and only if $\gamma$ has a positive real part. Thus, we
find that $\gamma$ and $\phi(s)$ must be solutions to the
generalised eigenvalue problem
\beq\label{eq:gen_eig_problem}
   \gamma\, \mathcal{L}[\phi(s)] = \mathcal{A}[\phi(s)].
\eeq
Note that this is equivalent to solving the eigenvalue equation
$\frac{\partial}{\partial t} \delta u_s(s) = \gamma \, \delta
u_s(s)$.

In order to find good (numerical) approximations for the
eigenvalues and eigenvectors, we again expand the linear operators
$\mathcal{L}$ and $\mathcal{A}$ in the $n+1$ first shifted
Legendre polynomials, so the eigenvector $\phi(s)$ is determined
by the coefficients $a_i$ in the expansion $\phi(s) = \sum_{i=0}^n
a_i\,\bar{P}_i(s)$. See Appendix~\ref{app:Dynamic stability
details} for details of how to solve for the coefficients $a_i$
numerically.

\begin{figure}
   \centerline{\includegraphics{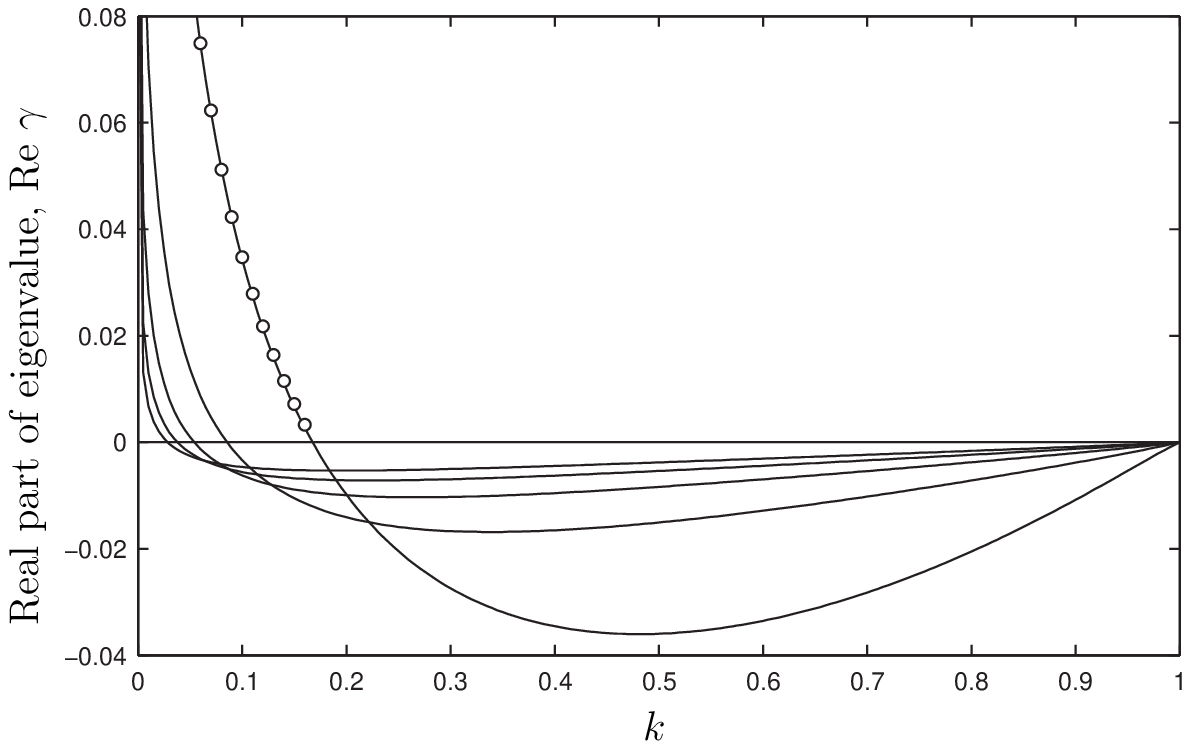}}
\caption{\label{fig:real_part_eig} The real part of the 11 largest
eigenvalues (ordered by modulus) as a function of the consolation
prize $k$, from the expansion in the 51 first Legendre
polynomials. We also show the growth rate for a small
perturbation, estimated from simulations (white circles) for some
values of $k$.
}
\end{figure}

The perturbation corresponding to an eigenvector $\phi(s)$ grows
at a rate given by the real part of the corresponding eigenvalue,
$\textrm{Re }\gamma$. The growth rate for the different
eigenvectors are shown in Fig.~\ref{fig:real_part_eig}, as a
function of the consolation prize $k$. For $k^*_\text{dyn} < k <
1$, where $k^*_\text{dyn} \approx 0.1675$, there is no eigenvalue
with positive real part. The value of $k^*_\text{dyn}$ does not
seem to be sensitive to the number of Legendre polynomials in the
expansion, when $n$ is large enough ($n \gtrsim 20$). Thus, in
this region all sufficiently small perturbations will decay
asymptotically, and the stationary distribution $u_s(s)$ is
dynamically stable. As a consequence of this, the population will
resist invasion of any mixed strategy, except the strategy $p(x) =
\lambda\,e^{-\lambda\,x}$, where $x$ is the waiting-time and
$\lambda$ is given by Theorem~\ref{thm:distribution}. This
strategy mimics a randomly chosen available player in the
stationary distribution, so all players get the same fitness
values as before. We also show an estimate of the growth rate from
simulations (see the circles in Fig.~\ref{fig:real_part_eig}), and
find good agreement with the growth rate predicted by the leading
eigenvalue, in the region where the growth rate is positive.

% fig 4
%======================================================================
%======================================================================

\begin{figure}
   \centerline{\includegraphics{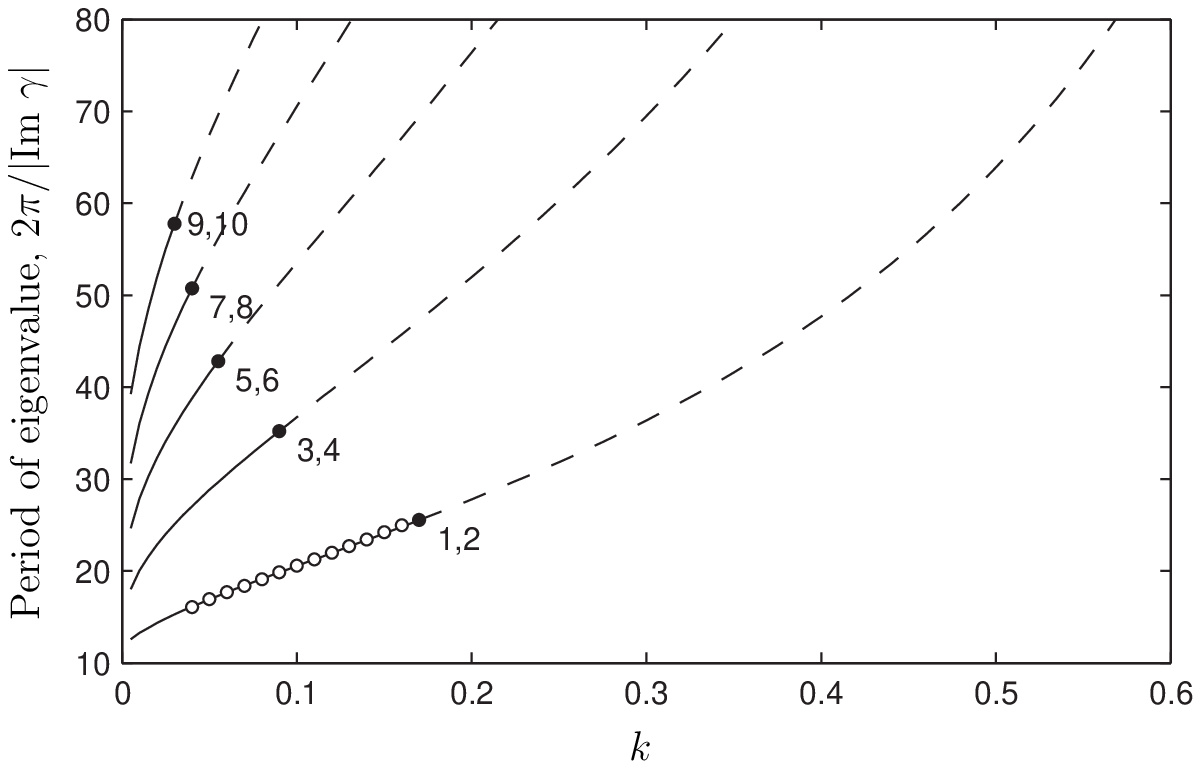}}
\caption{\label{fig:period_eig}  The period $2\pi/|\textrm{Im
}\gamma|$ as a function of the consolation prize $k$, where $\textrm{Im }\gamma$ is
the imaginary part of the eigenvalue $\gamma$. For each
eigenvalue, the curve is drawn solid where the real part is
positive, and is drawn dashed where it is negative. The black dots
shows the points of transition from negative to positive real
parts, and the number by the dots is the number of the eigenvalue
when sorted by decreasing magnitude. We also show the period of a
small perturbation, estimated from simulations (white circles) for
some values of $k$.}
\end{figure}

In general, a perturbation does not only grow or decay but also
oscillates with characteristic frequencies. In
Fig.~\ref{fig:period_eig} we show the period of these
oscillations, given by $2\pi/|\textrm{Im
}\gamma|$, as a function of $k$. The black points indicate where
the real part of the eigenvalue becomes positive (see
Fig.~\ref{fig:real_part_eig}). Estimated values from simulations
are shown as white circles, and we find good agreement with the
period rate predicted by the leading eigenvalue.

% fig 5
%======================================================================
%======================================================================

When $\mu(\pavail) \propto \pavail^{-1}$, the calculations of the
dynamical stability gives $k^*_\text{dyn} \approx 0.1197$, when $n
= 50$. We find as before, that this limit converges quickly for
$n$ large enough (i.e. $n \gtrsim 20$).

% --------------------------------------------------------------
% --------------------------------------------------------------
% --------------------------------------------------------------
% --------------------------------------------------------------

\section{Long-term evolution as a function of the initial distribution}
\label{sec:global_convergence}

\begin{figure}
   \centerline{\includegraphics{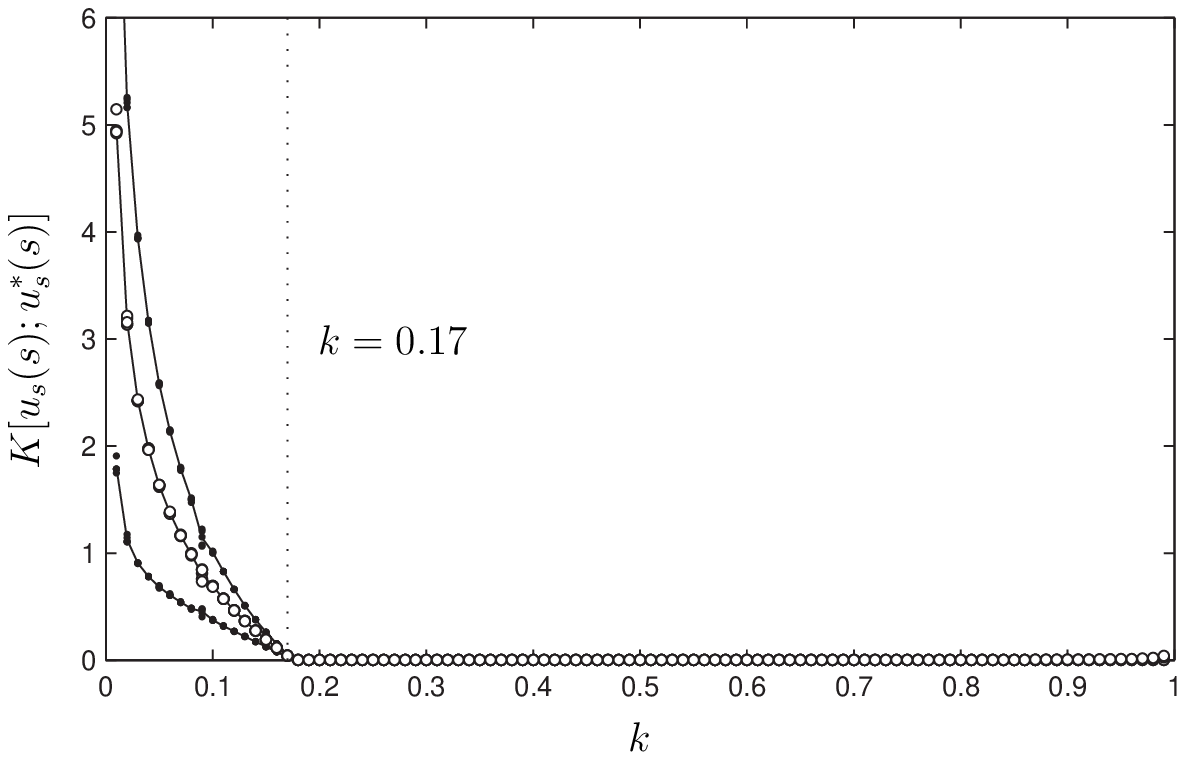}}
\caption{\label{fig:kullback_dist}
The asymptotic time-averaged Kullback distance
$K[u_s(s);u^*_s(s)]$ from the stationary distribution $u^*_s(s)$
(see (\ref{eq:losning_stat})) to the distribution $u_s(s)$, from
five different initial distributions (white circles). The points
corresponding to one standard deviation up and down is shown as
black dots, with one dot for each initial distribution. Solid
lines are averages over all initial distributions. The dotted
vertical line show $k = 0.17$, denoting the point where the
stationary distribution becomes stable. Convergence to the
stationary distribution is slow for $k$ close to one, since the
fitness differences are small. Note that the results are almost
indistinguishable for the five different initial distributions.
}
\end{figure}

In order to assess the global convergence of the dynamics to the
stationary distribution (c.f. Fig.~\ref{fig:time_evol}), we have
compared the asymptotic evolution of $u_s(s)$ when started from
different initial distributions:
   $u_s(s) = 1$,
   $u_s(s) = 2(1-s)$,
   $u_s(s) = 2(1+s)/3$,
   $u_s(s) = 2s$, and
   $u_s(s) = 6s(1-s)$.
In the simulations, the continuous parameter space $[0,1]$ was
approximated by $1,000$ equally spaced atoms in the interval
$(0,1)$, and time was divided into discrete steps of $0.1$ units
of time. The first $20,000$ units of time was considered
transient, and was discarded. The deviation from the stationary
distribution $u^*_s(s) = 2 \left[ 1 - (1-k)s \right]/(1+k)$ was
then measured at $100$ times, $10$ units of time apart, by the
Kullback distance \citep{kullback59}
\[
   K[u_s(s);u^*_s(s)] = \int_0^1 u^*_s(s) \ln \frac{u^*_s(s)}{u_s(s)} \, ds.
\]
The Kullback distance is always non-negative, and since $u^*_s(s)
\ge 0$ for all $s$, it is zero if and only if $u_s(s) = u^*_s(s)$
for all $s$ \citep{kullback59}. Second, the Kullback distance is
invariant under a change of variables, so it does not matter which
parameter (e.g. $x$ or $s$) we use to characterise a strategy.
From the samples we estimate the time average and standard
deviation of the distance. In Fig.~\ref{fig:kullback_dist} we show
the asymptotic average Kullback distance, plus and minus one
standard deviation, as a function of $k$ for the different initial
distributions above.

% fig 6
%======================================================================
%======================================================================

We summarise the results from these experiments as follows: First,
for $k > 0.17$ we find that the distribution of waiting-times
converges to the stationary distribution for the initial
distributions above. Second, for $k < 0.17$ we find that the
average of the asymptotic distance, and also its standard
deviation, is almost identical for the initial distributions
above. Inspection of the time evolution of the distribution gives
that the dynamics seems to converge to a limit cycle (c.f.\@
Fig.~\ref{fig:time_evol} for an example). This indicates that the
dynamics converge to stable limit cycle for these values of $k$.
Third, since the initial distributions was chosen to be very
different (some dominated by large waiting-times, some by low
waiting-times and other intermediate) we conjecture that these
results hold for any initial distribution with full support.

% --------------------------------------------------------------
% --------------------------------------------------------------
% --------------------------------------------------------------
% --------------------------------------------------------------

\section{Explicit time cost}
  \label{sec:explicit}

We return to models of an explicit cost for the duration of a
contest. Let $S(x)$ be the expected score per game for a player
with waiting-time $x$ (the cost not included), and let $D(x)$ be
expected duration of a contest. Suppose the cost per unit of time
during the contest is $c$. The expected fitness is then given by
\[
   f(x) = N_\text{G}(x) \left[ S(x) - c\,D(x) \right].
\]
Since by (\ref{eq:N_G(x)}), $N_\text{G}(x) = [\,1/\nu +
D(x)\,]^{-1}$ where $\nu = 2\,\pavail\,\mu(\pavail)$, we have
\[
   f(x) = N_\text{G}(x)\,S(x) - c + c\,N_\text{G}(x)/\nu.
\]
Note that since $N_\text{G}(x)$ is decreasing with the
waiting-time, players with small waiting-time are less influenced
by a high cost $c$, compared to players with higher waiting-time.

The average fitness in the population is by (\ref{eq:u(x)}) and
(\ref{eq:fbar}):
\be
   \bar{f} &=& \int_0^\infty \frac{\pavail\,\nu\,\rho(x)}{N_\text{G}(x)} \left[ N_\text{G}(x)\,S(x) - c + c\,N_\text{G}(x)/\nu \right] \,dx \nonumber\\
   &=& \pavail\,\nu\,\frac{1+k}{2} - c + c\,\pavail = (1+k)\,\pavail^2\,\mu(\pavail) - c\,(1 - \pavail). \nonumber
\ee
Thus, the average fitness is lowered by a factor $c\,(1 -
\pavail)$ by the introduction of the cost $c$.

We derive the stationary distribution with positive cost $c$ from
\[
   S(x) - c\,D(x) = \bar{f}\,[\, 1/\nu + D(x) \,]
\]
by letting $f(x)=\bar{f}$ as in Section~\ref{sec:stationary
distributions}. We find that for any cost $c > 0$, there is a
unique stationary distribution $u(x)$ involving all strategies
$x$, given by
\[
   u(x) = (2 - \pavail)\,\lambda\,e^{- \lambda\, x} - (1-\pavail)\,2\lambda\, e^{- 2\lambda \, x}
\]
where $\lambda = \pavail^2\,\mu(\pavail)/(1 - \pavail)$. The
fraction of available players $\pavail$ is the solution to
\beq\label{eq:pavial_cost}
   2\,k\,\pavail\,\mu(\pavail) - (1+k)\,\pavail^2\,\mu(\pavail) + c\,(1 - \pavail) = 0.
\eeq
The distribution of strategies among the available players is
$\rho(x) = \lambda\,e^{-\lambda\, x}$.

Note that since $\mu(\pavail)$ is assumed to be bounded, $\pavail
\rightarrow 1$ as $c \rightarrow \infty$ in order to fulfill
(\ref{eq:pavial_cost}). In the limit of $c = 0$ we recover
$\pavail = 2k/(1+k)$, and for $c > 0 $ we have $\pavail >
2k/(1+k)$. We now give the solution to (\ref{eq:pavial_cost}) for
two cases of special interest: for constant $\mu(\pavail)$ and for
$\mu(\pavail) \propto \pavail^{-1}$, we have
\[
   \pavail  = \frac{2\,k - c + \sqrt{(2\,k - c)^2 + 4\,(1+k)\,c}}{2\,(1+k)} \text{ and } \pavail = \frac{2\,k + c}{1 + k + c} \text{, respectively.}
\]
We will now show that one can find a $\tilde{k}$, such that the
fitness of the strategy with explicit time cost $c$ and the
consolation prize $k$, equals the fitness of the strategy with no
explicit cost and  the consolation prize $\tilde{k}$.

With the probability of winning a game, $P(x) = \int_0^x
\rho(y)\,dy$, we have the expected score $S(x) = k + (1-k)\,P(x)$,
and thus the fitness $f(x)$ may be written
\[
   f(x) = \left( \frac{c}{\nu} + k \right)\! N_\text{G}(x) + (1-k)\,N_\text{G}(x)\,P(x) - c .
\]
We compare this expression to the fitness of the strategy when $k
= \tilde{k}$ and $c = 0$:
\[
   f_{c\,=\,0}(x) = \tilde{k} \, N_\text{G}(x) + (1-\tilde{k})\,N_\text{G}(x)\,P(x).
\]
Since the growth rate of a strategy is proportional to the
difference between the fitness of the strategy and the average
fitness in the population, it does not matter if we add a number
to all fitness values and multiply them with a positive number.
Note that these numbers may change with time, as long as they are
the same for all strategies. Thus, we see that if we set
\[
   f_{c\,=\,0}(x)/(1 - \tilde{k}) = (f(x) + c)/(1 - k)
\]
we get equality for all $x$ if and only if
\beq\label{eq:c to 0 mapping}
   \frac{\tilde{k}}{1 - \tilde{k}} =  \frac{c/\nu + k}{1 - k} \quad \Leftrightarrow \quad \tilde{k} = 1 - \frac{1 - k}{1 + c/\nu}.
\eeq
Note that $c/(c + \nu) \le  \tilde{k} \le 1$ for all $k$, and
especially $\tilde{k} > k$ when $c > 0$.

When the expected time between games, $\nu$, is independent of the
fraction of available players in the population, $\mu(\pavail)
\propto \pavail^{-1}$, then $c/\nu$ is constant in time. It
follows that the dynamical equations are exactly the same for both
sets of parameters, apart from a scaling of the evolutionary time.

When $\nu$ depends on $\pavail$, we cannot map the dynamics for a
given combination of $k$ and $c$ onto $k = \tilde{k}$ and $c = 0$,
since the equivalent consolation price $\tilde{k}$ would then
change with time (unless the population is stationary). However,
since $\tilde{k}$ increases with decreasing $\nu$ according to
(\ref{eq:c to 0 mapping}), we expect that the domain of stability
is increased significantly.

% --------------------------------------------------------------
% --------------------------------------------------------------
% --------------------------------------------------------------
% --------------------------------------------------------------
\section{Comparison to other models}
\label{sec:comparison to other models}

In this section, we discuss how our results pertain to the models
of \citet{cannings_whittaker94} and \citet{hines77}.

\subsection{The model of Hines}

In Hines' model animals forage for food, and the events of finding
food are assumed to be a Poisson process with rate $\alpha$ per
animal. Each food parcel corresponds to one fitness unit.
Sometimes the animal gets to consume the food parcel without
challenge, but with probability $\pi$ the animal is challenged and
enters a war of attrition for the food parcel. The time spent
foraging and in contests are assumed to be statistically
independent. During foraging, the animal has energy consumption
$\mu_\text{F}$ fitness units per unit of time, and during a
contest the corresponding energy consumption is $\mu_\text{C}$
fitness units per unit of time.

Since the evolutionary dynamics modelled by replicator dynamics is
invariant when the fitness undergo addition of constants and
multiplication by positive constants, we find a relation between
the the consolation prize $k$ used in our model, in the absence of
explicit time cost, and the parameters in Hines' model:
\beq\label{eq:k hines}
   k = \frac{\alpha(1 - \pi) + \mu_\text{C} - \mu_\text{F}}{\alpha + \mu_\text{C} - \mu_\text{F}} .
\eeq
Note that the right hand side is positive for realistic choices of
the parameters. Naturally, $\pi = 0$ corresponds to $k = 1$, since
any food parcel found is uncontested. With $\mu_\text{C} =
\mu_\text{F}$, the relation is simply $k = 1 - \pi$.

The relation (\ref{eq:k hines}) completely maps the model by Hines
onto our model. Thus, the whole range of behaviours of Hines'
model is captured by the single value of $k$. Specifically, we
note that the stationary distribution characterising our model and
the corresponding stability analysis in Section~\ref{sec:dynamic
stability} apply.

\subsection{The model of Cannings and Whittaker}

In \citep{cannings_whittaker95} the authors studied a model where
players repeatedly meet each other in the war of attrition game,
with a score of zero for the loser, i.e. $k=0$, and with an
explicit time cost $c$. The waiting-times are restricted to the
set of positive integers, and the players wait one unit of time
between games. In their model, as in Hines' and ours, it is
assumed that players finishing faster get to play more frequently.
In addition to this they also introduced \emph{discounting}, where
scores received earlier in a sequence of games are valued higher.

We apply our method as described in Section~\ref{sec:social
dynamics model} to analyse the model. We find: Initial
correlations between players decay as the evaluation period
proceeds. Hence, in the limit of long evaluation periods, and when
the discount rate is sufficiently low, the average fitness over
the evaluation period, for a player, equals the expected fitness
in the equilibrium distribution. The time between games $\nu = 1$
gives $\mu(\pavail) = (2\pavail)^{-1}$. Since $\nu$ is constant,
we can map the evolutionary dynamics onto the dynamics in the
absence of explicit time cost and with consolation prize $k = c/(1
+ c)$. In Appendix~\ref{sec:discrete_waiting-times_stat_distr} we
derive the unique stationary distribution (involving all
waiting-times), for this model. With $\rho_i$ as the fraction of
the available players with waiting-time $i$, and with $u_i$ as the
corresponding fraction of the population, we find
\[
   \rho_1 = 4\,a - \frac{2\,c}{1+2\,c},\qquad u_1 = \frac{8\,a \left( 2\,a - c \right) }{1 + 2\,c},
\]
where $a = ( 3 + 6\,c - \sqrt{9 + 4\,c + 4\,c^2})/8$. For $i \ge
2$, we find
\be
   \rho_i &=& \left( 4\,a - 1 - 2\,c \right) \! \left( {\sqrt{1 + a^2}} - 1 - a \right) \! {\left( {\sqrt{1 + a^2}} - a \right) }^{i - 2} \quad\text{and} \nonumber\\
   u_i &=& \frac{2\,a\,\rho_i
     \left[ a + 2\,a\,c - ( 1 + a)\rho_i -
       ( \sqrt{1 + a^2} - 1)( 5 + 2\,c + \rho_i)  \right] }{
     {\left( 1 + a - \sqrt{1 + a^2} \right)}^2 \left( 1 + 2\,c \right) }.
\ee
Numerical simulations show that the dynamics converge to the
stationary distribution when $c \gtrsim 0.13$. For smaller costs,
the dynamics seems to converge to a stable limit cycle. Thus, the
dynamical properties of this model are similar to the dynamical
properties of our model where $\mu(\pavail) \propto \pavail^{-1}$,
and where the waiting-time may take any positive value.

% --------------------------------------------------------------
% --------------------------------------------------------------
% --------------------------------------------------------------
% --------------------------------------------------------------

\section{Discussion and concluding remarks} \label{sec:discussion}

We study the evolutionary dynamics of a population, where
individuals interact according to a social dynamics. In particular
we study a variation of the war of attrition (or the waiting-game)
in which the explicit time cost is replaced by an implicit cost,
due to less frequent game participation for those involved in
longer contests. Players are characterised by the waiting-time,
the time they are prepared to wait before they give up to the
opponent. The player with the larger waiting-time gets the score
$1$, while the other player gets the consolation price $k \in
[0,1]$. Players are in one of two states: either they are engaged
in a game or they are available for entering a new game. We
analyse the time evolution of the distribution of strategies in
the population when the population size is fixed, and players
produce offspring that survive to the next generation in
proportion to their fitness.

The fitness of a player is a product of two factors: the first is
the expected score per game, determined by the probability of
winning against a randomly chosen player, and the second is the
number of games played per unit time. There are two ways a player
may attempt to increase its fitness: it can either try to increase
the probability of winning the games played (by increasing the
waiting-time), or to increase the expected number of games (by
decreasing the waiting-time). In general, it is not possible for a
player to have both a high chance of winning a game, and to play
many games per unit of time. Thus, the second factor reflects an
implicit time cost.

We summarise our results as follows: First, the average fitness in
the population is determined by the fraction of the population
that is available, or equivalently, by the average duration of
games.

Second, when players follow deterministic strategies there is, for
all $k$, a unique stationary population involving all
waiting-times. When the consolation prize $k$ approaches the score
of 1 for winning a game, we find that the stationary distribution
is characterised by strategies with small waiting-times. For small
$k$, the stationary distribution is characterised by large
waiting-times: as $k$ approaches zero, the average waiting-time in
the stationary population diverges.

Third, the stability of the stationary state depends on the
consolation prize $k$. The stationary state is evolutionarily
stable for large enough values, $k \gtrsim 0.5196$ for constant
$\mu$, and $k \gtrsim 0.6738$ for $\mu \propto \pavail^{-1}$.
Below this level the evolutionary stability is lost, but the
system exhibits a weaker form of stability, dynamic stability,
when $k \gtrsim 0.1675$ for constant $\mu$, and when $k \gtrsim
0.1197$ for $\mu \propto \pavail^{-1}$. Here perturbations may
lead to a transient oscillatory behaviour, after which the system
is brought back to the stationary state again. Although the
perturbations eventually vanish, the transient period may be very
long. For small $k$ the stationary state is unstable, showing an
oscillatory behaviour in the average waiting-time. It shall be
noted that, in this model evolutionary stability does not imply
dynamical stability.

Fourth, the numerical simulations indicate that the stationary
distribution is a global attractor when $k$ is in the dynamically
stable region. When the stationary state is not dynamically
stable, the dynamics seems to converge to a stable limit cycle
from any initial distribution. The period and shape of the limit
cycle depends on $k$.

Fifth, there is an affine transformation of the fitness values in
a population, from a population with positive time cost $c$ to a
population with no explicit time cost, where the new consolation
price $\tilde{k}$ is now a function of the explicit time cost, the
original consolation price $k$, and the expected time between
games in the original population. When the time between games is
independent of the fraction of available players in the
population, the time evolution of the two populations are thus the
same.

Sixth, the implicit time cost is increasing with the explicit time
cost $c$. Thus, players with small waiting-time have an advantage
in that they are less affected by this cost.

Seventh, the models of \citet{hines77} and
\citet{cannings_whittaker94} both correspond to a time between
games independent from the population composition, hence our
methods and results apply to these models, when the number of
games per generation is large.

The oscillatory behaviour of the dynamics for small $k$ can be
intuitively understood by the following simplified arguments: if
the consolation price $k$ is small, and the average expected
waiting-time for each game is low, the players gradually increase
their waiting-time to win more games, driving the maximum
waiting-time up again as in an arms race. Here, the contribution
from the increase in the score per game is more important than the
decrease in the number of games that follows. Eventually, when the
average expected waiting-time becomes high enough, a small
fraction of the population can gain profit by choosing a low
waiting-time. The lower the waiting-time, the sooner the switch
becomes advantageous. Now, the implicit time cost due to long
games is so high that it is better to play many games but to win
few of them. As more players switch to short waiting-times, the
average expected waiting-time decreases and at some stage an arms
race is started. This creates the observed patterns of
oscillations (see Fig.~\ref{fig:time_evol}).

The domain of dynamical and evolutionary stability was found by
expanding a perturbation of the stationary distribution in a
suitable function basis to a finite size, and then evaluating the
eigenvalues for this approximation numerically. This proves only
that the distribution is unstable when we can identify a
perturbation that grows. It remains to be rigorously proven that
the distribution is stable in these domains, and to calculate the
exact values of $k$ where the distribution goes from stable to
unstable as $k$ decreases. Another open question is whether there
are eigenvalues in (\ref{eq:gen_eig_problem}) with a magnitude
arbitrarily close to zero. If so, there are perturbations that
decay arbitrarily slowly, and it is then necessary to analyse the
model also in the situation where perturbations are frequent
compared to the relaxation time of the perturbations. This can be
modelled by modifying the replicator dynamics to take into account
the rates by which mutations transform one strategy into another.

The players with zero waiting-time play a central role in
determining the shape of the stationary distribution. They never
win a game unless they play against another player with
waiting-time zero, but neither are they subject to the
duration-dependent costs: they play the maximum number of games
per unit of time, and they have no explicit time-cost. The fitness
of these players depends only on the fraction of available players
in the population, as does the average fitness in the population,
and one can thus say that the fitness at the stationary
distribution is determined by the players with waiting-time zero.

The approach presented with social dynamics giving rise to
non-trivial dependence of the duration of the game can be useful
for studies of game-theoretic problems in general. One example
could be the study of the Prisoner's Dilemma game with refusal in
which a player may quit a repeated game when encountering a
deviation from cooperation. In that case the role of the duration
could be the opposite, since it would be an advantage being
engaged with cooperative players for a longer time.

We conclude with a remark on empirical testing of the model. In
the literature, field surveys of the duration of contests are
compared to theory in order to assess the validity of the model
\citep[see e.g.][]{parker_thompson80}. In our model, the duration
of contests in the stationary population is exponentially
distributed, as it is in the classical war of attrition. Hence, we
conclude that this observable cannot be used to distinguish
between the two models. In order to recover the distribution of
waiting-times in the population, it is necessary to perform
experiments where the pairing process is controlled. If the
implicit time-cost is prominent in the population, we expect an
under-representation of players with short waiting-time in the
population, compared to what is expected from an exponential
distribution.

% --------------------------------------------------------------
% --------------------------------------------------------------
% --------------------------------------------------------------
% --------------------------------------------------------------

\section*{Acknowledgements}

We thank the anonymous referee for advice and comments.

% --------------------------------------------------------------
% --------------------------------------------------------------
% --------------------------------------------------------------
% --------------------------------------------------------------

%\bibliography{../wat_refs}

% --------------------------------------------------------------
% --------------------------------------------------------------
% --------------------------------------------------------------
% --------------------------------------------------------------

{
\appendix
\section{Dynamic stability evaluation details}
\label{app:Dynamic stability details}

We show how to represent the linearised dynamics in terms of two
matrices, corresponding to the linear operators $\mathcal{L}$ and
$\mathcal{A}$, and derive explicit expressions for the elements of
these matrices for the linearisation of the stationary state.

With the infinitesimal perturbation $\delta g(s) = \epsilon(s)$ of
the state $g(s)$, we give explicit expressions for the operators
$\mathcal{L}$ and $\mathcal{A}$:
\be
   \mathcal{L}[\epsilon(s)]
      &=& h(s)\,\epsilon(s) - (1-k)\, g(s) \int_0^1 \max(\ln s, \ln t)\, \epsilon(t)\,dt \nonumber\\
   \mathcal{A}[\epsilon(s)] &=&
         m(s)\,\epsilon(s) +  g(s) \int_0^1 dt\, \hat{S}(s,t)\, \epsilon(t)) \ - \nonumber\\
         && \hspace{2cm} -\ \hat{g}^2 \mathcal{L}[\epsilon(s)] - 2\, \hat{g}\, g(s)\, h(s) \int_0^1 \epsilon(t)\,dt.
\ee
where $\hat{S}(s,t) = \theta(t-s) + k\,\theta(s-t)$. We also
identify the adjoint operators, with respect to the inner product
$\int_0^1 v(s)\,w(s)\,ds$ of $v(s)$ and $w(s)$:
\be
   \mathcal{L}^\dagger [\epsilon(s)]
      &=& h(s)\,\epsilon(s) - (1-k) \int_0^1 g(t)\, \max(\ln s, \ln t)\, \epsilon(t) \,dt \nonumber\\
   \mathcal{A}^\dagger[\epsilon(s)]
      &=&  m(s)\,\epsilon(s) + \int_0^1  g(t)\, \hat{S}(t,s)\, \epsilon(t)) \,dt \ -  \nonumber\\
            && \hspace{2cm} -\ \hat{g}^2 \mathcal{L}^\dagger[\epsilon(s)] - 2\, \hat{g} \int_0^1 g(t)\, h(t)\,\epsilon(t)\,dt.
\ee
At the stationary distribution, $\mathcal{L}$ is self-adjoint but
$\mathcal{A}$ is not. If we apply the linear operators
$\mathcal{L}$, $\mathcal{A}$ and $\mathcal{A}^\dagger$ to powers
of $s$, we get
\be
   \mathcal{L}[s^n] &=& \frac{1-k}{(n+1)^2} + s^n - (1-k)\left[ 1 + (n+1)^{-2} \right]s^{n+1}  \nonumber\\
   \mathcal{A}[s^n] &=& - \frac{2 + n - k}{(n+1)^2} + 2\, \frac{1-k}{n+1}\,s - n \frac{1-k}{(n+1)^2}\,s^{n+1} \nonumber\\
   \mathcal{A}^\dagger[s^n] &=& - \frac{4 + 3\,n + k\,(n^2 - 2)}{(n+1)^2(n + 2)} + \frac{(1-k)(n+2)}{(n+1)^2}\,s^{n+1} ,
\ee
so polynomials over the interval $[0,1]$ are suitable as a basis
for expanding $\mathcal{L}$ and $\mathcal{A}$. We choose the
shifted Legendre polynomials $\bar{P}_i = P_i(2s-1)$, with the
property \citep{abramowitz_stegun72}:
\be\label{eq:legendre_norm}
   \int_0^1 \bar{P}_i(s) \, \bar{P}_j(s) \,ds
   \ =\ \frac{1}{2} \int_{-1}^1  P_i(x) \, P_j(x) \,dx
   \ =\ \frac{\delta_{ij}}{1+2i}.
\ee
We now insert the expansion $\phi(s) = \sum_{i=0}^n
a_i\,\bar{P}_i(s)$ in the eigenvalue relation
(\ref{eq:gen_eig_problem}):
\[
  \gamma\,\mathcal{L}[\,\sum_{j=0}^{n} a_j\,\bar{P}_j(s)\,] = \mathcal{A}[\,\sum_{j=0}^{n} a_j\,\bar{P}_j(s)\,].
\]
Since the set of Legendre polynomials form a basis, the projection
on all $\bar{P}_i(s)$ must be zero, we have that for all $i \ge
0$,
\[
   \gamma \int_0^1 \bar{P}_i(s) \sum_{j=0}^{n} a_j\,\mathcal{L}[\,\bar{P}_j(s)\,] \,ds =
   \int_0^1 \bar{P}_i(s) \sum_{j=0}^{n} a_j\,\mathcal{L}[\,\bar{P}_j(s)\,] \,ds ,
\]
which we may write as
\[
   \gamma \sum_{j=0}^{n} a_j  \int_0^1 ds\, \bar{P}_i(s) \, \mathcal{L}[\,\bar{P}_j(s)\,] =
   \sum_{j=0}^{n} a_j \int_0^1 ds\, \bar{P}_i(s) \,\mathcal{L}[\,\bar{P}_j(s)\,].
\]
By defining matrices $L$ and $A$ with elements
\be
   L_{ij} &=& \int_0^1 \bar{P}_i(s) \, \mathcal{L}[\bar{P}_j(s)] \,ds \quad \text{and} \nonumber\\
   A_{ij} &=& \int_0^1 \bar{P}_i(s) \, \mathcal{A}[\bar{P}_j(s)] \,ds
          \ =\ \int_0^1 \mathcal{A}^\dagger[\bar{P}_i(s)] \, \bar{P}_j(s) \,ds .
\ee
we can express (\ref{eq:gen_eig_problem}) as an eigenvalue problem
in the coefficient vector $a$:
\[
   \gamma \sum_{j=0}^{n} L_{ij} \, a_j = \sum_{j=0}^{n} A_{ij} \, a_j.
\]
Note that due to the normalisation (\ref{eq:dg norm cond}), we
need to take $a_0 = (1-k)\,a_1/3$.

Since any polynomial of degree $n$ can be written as a linear
combination of Legendre polynomials $P_0(s) \ldots P_n(s)$, it
follows from (\ref{eq:legendre_norm}) that $\bar{P}_n(s)$ is
orthogonal to any polynomial with degree $n-1$ or less on the
interval $[0,1]$. From this follows that when $i > j+1$, $L_{ij}$
and $A_{ij}$ are zero since $\mathcal{L}[\bar{P}_j(s)]$ and
$\mathcal{A}[\bar{P}_j(s)]$ are polynomials of degree $j+1$. When
$j > i + 1$, we have that $L_{ji} = L_{ij} = 0$ since
$\mathcal{L}$ is self-adjoint. Since
$\mathcal{A}^\dagger[\bar{P}_i(s)]$ is a polynomial of degree
$i+1$, we see that $A_{ij} = 0$. To conclude, both $L_{ij}$ and
$A_{ij}$ are zero when $|i - j| > 1$.

The shifted Legendre polynomials obey the recurrence relation
\citep{abramowitz_stegun72}
\[
   (n+1)\bar{P}_{n+1}(s) - (1 + 2\,n)(2\,s - 1)\bar{P}_n(s) + n\,\bar{P}_{n-1}(s) = 0.
\]
In conjunction with the above mentioned properties of the shifted
Legendre polynomials, we used this recurrence relation to
calculate the elements of $L_{ij}$ and $A_{ij}$:
\[
   L_{ij} = \left\{ \begin{array}{ll}
      \displaystyle 1 & \text{when  $i = j = 0$} \\
      \displaystyle \frac{1}{1 + 2\,i} - \frac{i^2 + i - 1}{2\,i\,( 1 + i)( 1 + 2\,i) }\,(1-k) & \text{when  $i = j$ and $i \ge 1$} \\
      \displaystyle -\,\frac{1 + i^2}{2\,i\,(2\,i -1)(2\,i + 1) }\,(1-k) & \text{when  $j = i - 1$} \\
      \displaystyle -\,\frac{2 + 2\,i + i^2}{2\,(1 + i)(1 + 2\,i)(3 + 2\,i)}\,(1-k) & \text{when  $j = i + 1$} \\
      \displaystyle 0 & \text{otherwise}
   \end{array} \right.
\]
and
\[
   A_{ij} = \left\{ \begin{array}{ll}
      \displaystyle -1 & \text{when $i = j = 0$} \\
      \displaystyle \frac{1-k}{3} & \text{when  $i = 1$ and $j = 0$} \\
      \displaystyle -\frac{1-k}{2\,i\,(1 + i)(1 + 2\,i)}  & \text{when  $i = j$  and $i \ge 1$} \\
      \displaystyle -\frac{(i-1)\,(1-k)}{2\,i\,(2\,i - 1)(2\,i + 1) } & \text{when  $j = i - 1$ and $i \ge 2$} \\
      \displaystyle \frac{( 2 + i)\,(1 - k ) }{2\,(1 + i)( 1 + 2\,i)( 3 + 2\,i) } & \text{when  $j = i + 1$} \\
      \displaystyle 0 & \text{otherwise}
   \end{array} \right.
\]

% --------------------------------------------------------------
% --------------------------------------------------------------
% --------------------------------------------------------------
% --------------------------------------------------------------

\section{Stationary distribution for discrete waiting-times}
\label{sec:discrete_waiting-times_stat_distr}

Here, we derive the stationary distribution for a population with
deterministic strategies, $\nu = 1$ and with waiting-times
restricted to positive integers, in the absence of explicit
time-cost ($c = 0$). Note that since $\nu$ is constant, any
combination of consolation prize and explicit time cost may be
mapped to an effective consolation prize and no explicit time
cost, as shown in Section~\ref{sec:explicit}.

With $\phi_i = \sum_{j=1}^i \rho_j$, the expected duration of game
for a player with waiting-time $i$ is
   $D_i = \sum_{j=1}^{i-1} j \rho_j + i\,(1 - \phi_{i-1})$,
and the expected score for the player is
   $S_i = \phi_{i-1} + (1+k)\rho_i/2 + k\,(1 - \phi_i)$.

At the stationary solution, we have
   $S_i = \bar{f} + \bar{f}\,D_i$ where $\bar{f} = \bar{f} = (1+k)\pavail/2$.
This implies
   $S_{i+1} - S_i = \bar{f}\, (D_{i+1} - D_i)$, or equivalently, $\rho_i + \rho_{i+1} = 2\,\bar{f}\,( 1 - \phi_i)/(1 - k)$,
which may be written as
   $\phi_{i+1} + 2\,a\,\phi_i - \phi_{i-1} = 2\,a$, where $a = \bar{f}/(1 - k)$,
for $i \ge 2$. The unique solution is
\begin{multline*}
      \phi_i = \frac{1}{2\,{\sqrt{1 + a^2}}} {\left( -{\sqrt{1 + a^2}} - a \right)}^{i-1} \left[ 1 - a - {\sqrt{1 + a^2}} + \left( a + {\sqrt{1 + a^2}} \right) A - B \right]\\
      + \frac{1}{2\,{\sqrt{1 + a^2}}} {\left( {\sqrt{1 + a^2}} - a \right) }^{i-1} \left[ a - 1 - {\sqrt{1 + a^2}} + \left( {\sqrt{1 + a^2}} - a \right) A + B \right] + 1
\end{multline*}
for $i \ge 1$, where $A$ and $B$ are constants to be determined.
Since $\sqrt{1 + a^2} + a > 1$ for all $a > 0$, the factor
   ${( - \sqrt{1 + a^2} - a )}^{i-1}$
grows without bound, so we must take
   $B = 1 - a - \sqrt{1 + a^2} + ( a + \sqrt{1 + a^2} ) A$
in order to have $\phi_\infty = 1$. For $i = 1$, we find $\phi_1 =
A = \rho_1$. With $S_1 = \frac{1+k}{2}\,\rho_1 + k\,(1 - \rho_i)$
and $D_1 = 1$, the stationarity condition gives $\rho_1 = 4\,a -
2\,k/(1-k)$.

We must now solve for the value of $a$. A simple calculation gives
\begin{multline*}
   S_i - (1 - k)\,a\,D_i =\\
   \frac{
   1 - 2\,a^3 \left( 1 - k \right)  + a\,k + 2\,a^2\,k - {\sqrt{1 + a^2}} \left( 1 + 2\,a\,k - 2\,a^2 \left( 1 - k \right)  \right)
   }{
   \left( 1 + a - {\sqrt{1 + a^2}} \right) {\left( {\sqrt{1 + a^2}} - a \right) }^2
   }
\end{multline*}
for $i \ge 2$. Setting this to zero, we find that there is one
valid solution:
\[
   a = \frac{3}{8\,\left( 1 - k \right) }\left(1 + k - {\sqrt{1 - \frac{14}{9}\,k + k^2}} \right).
\]
Another solution corresponds to negative fitness values, which is
not possible here. With this solution, and using $u_i =
\pavail\,\rho_i\,( 1 + D_i )$ and $\pavail = 2\,(1 - k)\,a/(1+k)$,
we find
\[
   \rho_1 = 4\,a - \frac{2k}{1-k} \text{ and } u_1 = \frac{16\,\left( 1 - k \right) a^2 - 8\,a\,k}{1 + k},
\]
and for $i \ge 2 $ we have
\be
   \rho_i &=& \left( 4\,a - \frac{1 + k}{1 - k} \right) \left( {\sqrt{1 + a^2}} - 1 - a \right) {\left( {\sqrt{1 + a^2}} - a \right) }^{i - 2} \quad\text{and} \nonumber\\
   u_i &=&
   \frac{1 + k + \left( a + \sqrt{1 + a^2} \right) \left( 1 - 4\,a\,\left( 1 - k \right)  + k \right)}{1 + k} \, \rho_i \ \nonumber -\\
   &&\hspace{5cm}-\ \frac{2\,a\,\left( a + {\sqrt{1 + a^2}} \right) \left( 1 - k \right) }{{\left( 1 + a - {\sqrt{1 + a^2}} \right) }^2 \left( 1 + k \right) } \, \rho_i^2 .\nonumber
\ee
}

\end{document}